\documentstyle{mn}
\def\today{\ifcase\month\or
 January\or February\or March\or April\or May\or June\or
 July\or August\or September\or October\or November\or
 December\fi\space\number\day, \number\year}

\def\todmy{\number\day\space\ifcase\month\or
 January\or February\or March\or April\or May\or June\or
 July\or August\or September\or October\or November\or
 December\fi\space\number\year}

\newcommand{\bdisp} {\begin{displaymath}}
\newcommand{\edisp} {\end{displaymath}}
\newcommand{\beqn} {\begin{equation}}
\newcommand{\eeqn} {\end{equation}}
\newcommand{\beqr} {\begin{array}}
\newcommand{\eeqr} {\end{array}}

\newcommand{\tal}{\it et al. \rm}
\newcommand{\AAA}{{A\&A}}

\newcommand{\ApJ}{{ApJ}}
\newcommand{\ApJL}{{ApJ}}
\newcommand{\ApJS}{{ApJS}}
\newcommand{\AJ}{{AJ}}
\newcommand{\AR}{{ARA\&A}}

\newcommand{\MN}{{MNRAS}}

\newcommand{\PASJ}{{PASJ}}
\newcommand{\Nat}{{Nat}}

\title{Optimal softening for force calculations in collisionless
N-body simulations} 
\author[E. Athanassoula, E. Fady, J.C. Lambert, A. Bosma]
       {
       E. Athanassoula$^{1}$, E. Fady$^{1}$, J.C. Lambert$^{1}$, 
	and A. Bosma$^{1}$ \\
$^1$ Observatoire de Marseille, 
2 Place Le Verrier, 
F-13248 Marseille Cedex 4, France \\}

\date{Accepted .
      Received ;
      }

\pagerange{\pageref{firstpage}--\pageref{lastpage}}
\pubyear{1999}

\begin{document}

\maketitle

\label{firstpage} 
\begin{abstract}

In N-body simulations the force calculated between particles
representing a given mass distribution is usually softened, to diminish
the effect of graininess. In this paper we study the effect of such a
smoothing, with the aim of finding an optimal value of the softening
parameter. As already shown by Merritt (1996), for too small a
softening the estimates of the forces will be
too noisy, while for too large a softening the force estimates are
systematically misrepresented. In between there is an optimal
softening, for which the forces in the configuration approach best the
true forces. The value of this optimal softening depends both on the
mass distribution and on the number of particles used to represent
it. For higher number of particles the optimal softening is
smaller. More concentrated mass distributions necessitate smaller
softening, but the softened forces are never as good an approximation
of the true forces as for not centrally concentrated
configurations. We give good estimates of the optimal softening for
homogeneous spheres, Plummer spheres,
and Dehnen spheres. We also give a rough
estimate of this quantity for other mass distributions, based on the
harmonic mean distance to the $k$th neighbour ($k$ = 1, .., 12), the mean
being taken over all particles in the configuration. Comparing homogeneous
Ferrers ellipsoids of different 
shapes we show that the axial ratios do not influence the value of the
optimal softening. Finally we compare two different types of
softening, a spline softening (Hernquist \& Katz 1989) and a 
generalisation of the standard Plummer softening to higher values of
the exponent. We find that the spline softening fares roughly as
well as the higher powers of the power-law softening and both give
a better representation of the forces than the standard
Plummer softening. 

\end{abstract}


%\newpage
\section{Introduction}
\indent

$N$-body codes are often used to simulate the dynamical evolution of
galaxies and galaxy systems, 
even though the number of particles that present day computer hardware and 
software can handle is smaller, by several orders of magnitude, than
the number of stars in even a small
galaxy. Because of 
this the particles do not represent individual stars, but 
should be considered as Monte-Carlo 
realisations of the mass distribution in a galaxy. In such
simulations, when close encounters between individual particles are
of no relevance to the physical problem under consideration, the
gravitational force between 
two particles is smoothed, in order to
reduce the spurious two-body relaxation due to a number of particles
necessarily much smaller than the total number of stars in the 
system. Although a large softening will ensure a low
relaxation rate, we can not increase this value at will, since a high
value of the softening introduces other drawbacks and in particular
a bias to the gravitational force.
 
The question we will address in this paper is what value of the 
softening should be used in order for $N$ particles to represent best a 
given density distribution. 
In a recent paper Merritt (1996, hereafter M96)  
argues that too 
small a value for the smoothing will give too noisy estimates, while too 
large a value will give a systematic misrepresentation of the
force. M96 and Athanassoula \tal (1998; hereafter A+98) found that
this optimal value 
of the softening depends on the number of particles $N$ approximately 
as $N^{-0.3}$. In this 
paper we will extend previous work to other configurations, other 
$N$-body algorithms for calculating 
the force and other functional forms for the force approximation. 
The two $N$-body methods under consideration are briefly described in  
section \ref{sec:notation}, where we present also the
tools we will use for our comparisons. The optimal softening 
for the case of a Plummer sphere is discussed 
in section \ref{sec:plummer}, including comparisons of $MISE$ and
$MASE$, as well as comparisons of calculations with GRAPE-3 and
GRAPE-4. Section \ref{sec:twoplum} presents the case of two Plummer
spheres and section \ref{sec:homo_dehnen} two other density distributions, one
more and the other less concentrated than the Plummer sphere. Section
\ref{sec:ferrers} discusses the effect of triaxiality, using the Ferrers'
ellipsoid. Different types of softening are introduced in section
\ref{sec:other-soft} and a different way of calculating the force, using a
hierarchical octal tree, in section
\ref{sec:tree}. Finally our results are
summarised and discussed in section \ref{sec:discuss}.

\section{Methods}
\label{sec:notation}

\subsection{Force evaluation methods}

Several methods have been developed for calculating the  
forces in self-consistent $N$-body simulations (see e.g. the reviews 
by Sellwood 
1987 and Athanassoula 1993). They all have their advantages and disadvantages 
and which one should be chosen depends to a large extent on the problem to be 
solved. We will here briefly describe two methods not using
either grids or expansions, in which case the introduction of the
softening is more straightforward.

1) Direct summation. This method involves calculating the forces 
between 
every pair of particles and then summing up all contributions to the 
force on a given 
particle. It is straightforward, easy to program and easy to 
vectorise and parallelise. For a fixed 
number of particles it is by far the most accurate 
method, since its only approximation is the introduction of softening, and can 
be used without any restrictions on geometry. 
This method was used heavily in the seventies and early eighties, but was 
soon considered as 
a dead end because of its large claims in CPU time. Indeed the amount of time 
necessary for one force calculation scales as $N(N-1)$, where $N$ the
number of particles in the simulation. 
Thus, until recently, direct summation could not be used freely for 
simulations where a 
sufficiently high number of particles is necessary, as e.g.
simulations of disc galaxies.
Two major advances in computer hardware have changed the situation in
the past few years. 

The first is the advent of parallel machines, either SIMD or MIMD, 
which, given the simplicity of the communications involved in the
direct summation method, make it
possible to achieve relatively high performances with little 
software investment. A large number of such systems are actually in
use, from the powerful many-node CRAY T3E, to ``Beowulf'' clusters,
which have been recently implemented in many universities and
research centers.

The second is the realisation of 
GRAPE, a dedicated computer card which performs the force calculation
by direct summation 
and which can be coupled to a  
standard workstation allowing one to achieve at relatively low cost an
excellent 
CPU performance. A series of such GRAPE boards have been built by the group in 
Tokyo University, starting with GRAPE-1 and evolving steadily to GRAPE-5,
while new members of this family are under development. For a brief history
of this project see Makino \& Taiji (1998) and references
therein. Boards with even numbers  
have high accuracy arithmetic and can be used for collisional simulations, 
where close encounters play an important role in the evolution of the system,
as for globular clusters and planetesimals. Boards with odd numbers have 
a more limited precision arithmetic and can only be used for
collisionless systems. Thus GRAPE-3 uses 14 bits to represent the masses, 
20 bits for the positions and 56 bits for the forces. 
Nevertheless this was shown to be sufficient for simulations (A+98),
since the dominant source of error is the noise in the form of
two-body relaxation, while the effect of the error in the force
calculations is comparatively smaller
(Hernquist \tal 1993, Makino 1994). 

Two such systems have been used for the calculations presented in this
paper, namely the GRAPE-3AF and GRAPE-4 systems in Marseille
Observatory. The former consists of five GRAPE-3AF boards coupled via
an Sbus/VMEbus converter to a Sun Ultra 2/200, and is described in
detail in A+98. The latter consists of one GRAPE-4 processor board and
one control board  (Makino \tal 1997) linked to an Alpha 500/500
workstation via a PCI interface board (Kawai \tal 1997).
 
2) Treecodes. 
With the help of treecodes considerably more particles can be used in 
$N$-body simulations, while one of the main advantages of direct 
summation codes, namely that they can be applied to systems with any
geometry, is preserved. 
They stem from the simple idea that when a group of particles is
sufficiently distant from another particle and its spatial extent is small
with respect to the distance separating it from the particle, then this 
group can be considered as one entity, and only monopole and, in some cases,
quadrupole terms need be retained for the calculation of the force
exerted by this group on the  
particle. Whether a group of particles can be considered as
one entity, or whether it has to be further subdivided in subgroups is
determined by the tolerance, or opening angle, which determines the
precision of the force calculation. In this way one obtains a considerable
saving over the number of force calculations necessary in a direct summation
code, so that for tree-codes the necessary time increases with the
number of particles  
$N$ as $NlogN$ or as $N$. The version of the treecode most commonly used
in astronomy  
is the Barnes-Hut algorithm (1986) and in particular its vectorised
version, freely  available from Hernquist (1987). 

The treecode shares several of the advantages of direct summation, namely
that it can be applied to distributions with any geometry and large spatial 
and/or temporal density variations, while being considerably faster. It
has, however, also many drawbacks. It is more difficult to program, and its
vectorisation, and particularly parallelisation, present several
difficulties. 
A more worrisome drawback is the fact that Newton's third law of motion is 
not necessarily obeyed. Take the example of an isolated particle A and a 
particle B in a cluster of particles far from A. Then the force of A on B 
is correctly calculated, while A will only feel the effect of the cluster 
of particles as a whole. Even so the {\it total} force on A from the whole 
of the cluster is adequately calculated and therefore one can expect a 
correct evolution of the system. Other drawbacks are discussed by Salmon and 
Warren 
(1994), who discussed the relative merits of opening angle criteria (or 
``Multipole Acceptability Criteria", as they call them).

The implementation of such a code on GRAPE systems is possible,
provided one keeps the monopole term only, and neglects quadrupole and
higher order terms (Makino 1991, A+98). Such terms are nevertheless
possible to include using a method described by Kawai \& Makino
(1999).

\subsection{Notations and computing miscelanea}
\indent

The usual way of softening the Newtonian force exerted on a particle 
by another particle is to by replace in the calculation of the potential
the distance between the two particles, $r$, by
$\sqrt{r^2+\epsilon^2}$, where $\epsilon$ is the softening parameter.
In this way the acceleration (or force per unit mass) on
particle $i$ from $N-1$ other particles is written as:

\begin{equation}
{\bf F}_i = G  \sum_{j=1}^{N} \frac {m_j({\bf x}_j - {\bf x}_i)}
{(|{\bf x}_i - {\bf x}_j|^2 + \epsilon ^2)^{1.5}}
\label{eq:force_plum}
\end{equation}

\noindent
Here $G$ is the
gravitational constant which we hereafter take to be equal to unity. In
the following we will assume for simplicity that all particles have 
the same mass, equal to $m$.
Each term in eq. (1) is equal to the force felt by a point of unit mass
in a Plummer sphere of scale-length $\epsilon$, and  we shall thus
hereafter refer to this softening for brevity as the 
Plummer softening. It was initially introduced by Aarseth (1963) in
simulations of clusters of galaxies. Although this is the most
commonly used form of the 
softened Newtonian force, it is not the only one. Alternative types of
softening will be described in section \ref{sec:other-soft}.

Independent of the way softening is introduced, it is clear that 
too little smoothing leads to noise and too much of it to the modeling of 
a gravity which is far from Newtonian. In order to compare the 
calculated forces to the true ones M96 introduced the quantities $ISE$
(Integrated Square Error), $ASE$ (Average Square Error),  $MISE$ (Mean 
Integrated Square Error) and $MASE$ (Mean Average Square Error). We
will use them here also, since they provide a very useful way of
quantifying how well the force of a given configuration is
approximated. We will, however, modify them somewhat for our present
needs, e.g. by introducing a multiplicative constant to
make them dimensionless, so that it is possible to make comparisons
between models with different total mass and size. We briefly summarise
all the definitions below.

Let ${\bf F}_{true} ({\bf x}_i)$ be the true force from
a given mass distribution at
a point ${\bf x}_i$, and let $F_i$ be the force calculated at the same 
position from a given $N$-body realisation of the mass distribution and
using a given  
softening and method. Then the average deviation between the two
forces is given by

\begin{equation}
ASE=\frac{\cal C}{N} \sum_{i=1}^{N}|{\bf F}_i-{\bf F}_{true}({\bf x}_i)|^2
\label{eq:ASE}
\end{equation}

\noindent
where the summation is 
over the $N$ particles in the realisation. We can similarly introduce the 
``integrated square error", or

\begin{equation}
ISE=\frac{\cal C}{M} \int \rho ({\bf x}) |{\bf F}({\bf x})-{\bf F}_{true}({\bf x})|^2 d{\bf x}
\label{eq:ISE}
\end{equation}

\noindent
where $\rho ({\bf x})$ is the true density at point ${\bf x}$,
$M$ is the total mass in the system, ${\bf F}({\bf x})$ is the force at
position ${\bf x}$ calculated from the $N$-body realisation of the
mass distribution, and the 
integral is taken over a volume in space encompassing the
configuration. For spherically symmetric  
mass distributions using the $ISE$ rather than the $ASE$ brings a 
substantial gain in CPU time. Indeed the integration over the angles 
can be done analytically and one is left with a one-dimensional integration 
along a radius. We will sometimes refer to this as the radial $ISE$
algorithm. For this the gain in time is proportional to
$n_{int}/(N-1)$, where  
$n_{int}$ is the number of points used when calculating the integral. 
Nevertheless this is made at the expense of considerable noise, since
the number of points at which the integrand is calculated is
relatively small. This will be discussed further in section
\ref{sec:mise_mase}.  

In order to get rid of the dependence on the particular configuration, which 
is of no physical significance, we generate many realisations of the same 
smooth model and average our results over them. Thus the mean value of
the $ASE$ is 

\begin{equation}
MASE = \frac {\cal C} {N} <\sum_{i=1}^{N}|{\bf F}_i-{\bf F}_{true}({\bf x}_i)|^2>
\label{eq:MASE}
\end{equation}

\noindent
where $<>$ indicates an average over many realisations. Similarly for the 
mean value of the $ISE$ we get

\begin{equation}
MISE = \frac{\cal C}{M} <\int \rho ({\bf x}) |{\bf F}({\bf x})-{\bf F}_{true}({\bf x})|^2 d{\bf x}>
\label{eq:MISE}
\end{equation}

\noindent
Values of $\epsilon$ minimising the $MISE$ or the $MASE$ should give 
the optimal representation of the force of the system.

In the above ${\cal C}$ is a multiplicative constant, introduced 
to permit comparisons between different mass distributions. 
If we only want to assess the effect of the number of particles, 
as in M96 and sections \ref{sec:plummer} and \ref{sec:ferrers}, 
we can simply use ${\cal C}$ = 1.
On the other hand if more than one configuration are to be compared
then it may be necessary to use the multiplicative constant ${\cal C}$.
To make this clearer let us consider two Plummer spheres of the same
mass, and of which the second one has double the scale length of the
first one. In that case the optimal softening for the second should be
double the optimal softening for the first one. By rescaling the
second Plummer sphere, i.e. multiplying all distances by 0.5, we would
get for both the same softening, which 
is obvious since they are in fact the same object. It is thus
preferable to compare softenings after the objects have been rescaled
to the same mass and size. In a similar way if we want to compare the
appropriate softening for a Plummer sphere and that of a Dehnen sphere we
first have to rescale one of the two so that the two objects correspond
to the same mass and size, otherwise the comparison would not be
meaningful, and would only tell us that bigger objects need
bigger softenings. Once the objects are rescaled to the same mass and size,
then the comparison of the optimal softenings could tell us something
about the effect of the central concentration. Such a rescaling
is, however, not possible in all cases. If in a simulation we
want to see e.g. the evolution of a given Plummer sphere in presence
of a given Dehnen sphere, then we can not rescale each component
separately, without changing the problem we are considering. In this
case we have to search for the softening that would help best
represent the entire configuration, and then compare it with the
softening that represents best each of the two unscaled objects
individually. The above is of course only common sense. We have
nevertheless explained it in some detail here since otherwise it may
be unclear 
to the reader why, in the following sections, in some cases we use a
scaling and in others we do not.

We will use the following definition for ${\cal C}$ which makes the
$MISE$ and $MASE$ quantities dimensionless:

$${\cal C} = F^{-2} ({\cal R})$$

\noindent
or, equivalently,

$${\cal C} = {\cal R}^{4} M^{-2}$$

\noindent
Here $M$ is the total mass in the configuration, ${\cal R}$ is a
characteristic radius, e.g. the half mass radius, and $F({\cal R})$ is
the force at that radius. Similarly, in such cases, the softening
should also be scaled with the adopted characteristic radius,
since it has units of length.

It is debatable whether in the definition of $ISE$ it would have been
better to use the numerical or the analytical 
density of the configuration. 
We have chosen the latter for 
two reasons. One is continuity with the definition of M96, and the 
other is that the calculation of the density of an $N$-body configuration 
is not straightforward and may, by itself, introduce further
uncertainties and complications.
As will be shown in the next section, for sufficiently large number of 
realisations, the present definitions of $MISE$ and $MASE$ give the same 
results and this legitimises our choice for the density. 

In all the examples in this paper, unless otherwise stated, we have
used 6 $\times 10^6 / N$ realisations. 
Because of the high CPU speed of our two GRAPE systems we
have used, whenever possible, $MASE$ calculations on GRAPE. The two
notable exceptions are the calculations with alternative softening
(section~\ref{sec:other-soft}) and the calculations with the standard
treecode (section~\ref{sec:tree}),
both of which can not be done on GRAPE, and for which we have confined
ourselves to radial $MISE$ calculations.

For the treecode calculations on GRAPE we have used the software
described in detail in A+98. For the standard treecode calculated on a
workstation we have used 
the Barnes version of the Barnes-Hut 
algorithm (Barnes and Hut 1986) included in the NEMO package
(cf. Teuben 1995).
This version of the treecode can use up to
quadrupole terms.

For the integration along the radius in the radial $ISE$ calculation we use 
the alternative extended Simpson's rule (Press \tal 1988), which has 
an accuracy of ${\cal O}(N^{-4})$, and 100 points along the line of integration. 
For the calculation of the radial $ISE$ values for the Plummer sphere we have 
used as an upper limit of the integration $L=20a_p$, where $a_p$ is the 
scale-length of the Plummer sphere. This radius contains more than 99\% 
of the mass of the Plummer model, while the density has fallen to 
roughly $3 \times 10^{-7}$ of its central value. 

\section{Optimal smoothing for a Plummer sphere density distribution}
\label{sec:plummer}

\subsection{Dependence of the error on the softening and the number of
particles}
\label{sec:plummer_MASE}

\begin{figure}
\includegraphics{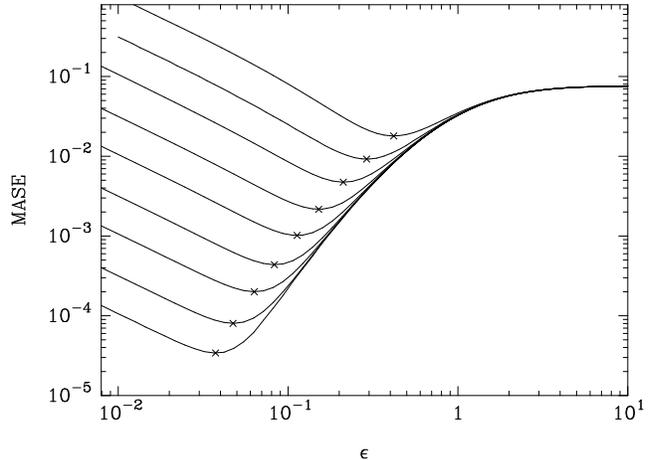}
\vspace{7.0cm}
\caption{
$MASE$ as a function of the softening $\epsilon$ for a Plummer
sphere. From top to bottom the 
curves correspond to 
$N = 30, 100, 300, 1~000, 3~000, 10~000, 30~000, 100~000, 300~000$, 
where $N$ is the
number of particles in the realisation of a Plummer sphere. The number
of realisations taken for the mean is $6 \times 10^6 / N$.
The position of the minimum error along a line corresponding to a
given $N$ is marked by
an $\times$, and the corresponding $\epsilon$ value is the optimal softening 
$\epsilon_{opt}$
for this number of particles. 
}
\label{repeat}
\end{figure}

Following M96 and A+98 we will use, as a first model, a truncated
Plummer sphere

\[\rho (r) =\left\{ \begin{array}{ll}
\frac {3 M_T} {4\pi a_p^3} (1 + r^2/a_p^2)^{-5/2} & \mbox{$r \leq R$} \\
0 & \mbox{$r > R$} \\
\end{array}
\right. \]

\noindent
where $M_T$ is the mass of the Plummer sphere had it extended to
infinity, $a_P$ is its scale-length and $R$ is its truncation radius. For
the remainder of this 
paper, unless otherwise noted, we will adopt, without loss of
generality, $a_P$~=~1, 
as a truncation radius the radius containing 0.999 $M_T$, and the mass
within the truncation radius, $M(\leq R)$ = 1.
The corresponding radial $MISE$ and $MASE$ values were calculated in M96,
for $N$ between 30 and 30~000, and in A+98 for $N$ between 30 and
300~000 with direct summation and 1~000~000 using a tree code. It was
found that, as expected, the error for a given number of particles
$N$, be it radial $MISE$ or 
$MASE$, shows a minimum for a given value of $\epsilon$. In
Figure~\ref{repeat} we show a number of examples for various values of
$N$. The information is essentially the same as in Figure 4 of
A+98, except now for $MASE$ rather than radial $MISE$.

For small values of the softening the noise
dominates the error. For this reason the $MASE$, for such values of
the softening, decreases steeply with $N$, the number of particles in
the configuration. Conversely for large values of the softening 
it is the bias that dominates. For a sufficiently large value of the
softening the $MASE$ does not show any dependence, either on the
softening, or on the number of particles. In this region the
inter-particle forces go as $r\epsilon^{-3}$, and tend to zero as
$\epsilon \rightarrow \infty$. Thus, for sufficiently large values of the
softening, the ${\bf F}_{true}({\bf x})$ term dominates in the
difference in equations (\ref{eq:ASE}), (\ref{eq:ISE}), (\ref{eq:MASE}) and
(\ref{eq:MISE}) and $MISE$ and $MASE$ tend to     

\begin{equation}
<\int \rho ({\bf x}) |{\bf F}_{true}({\bf x})|^2 d{\bf x}>
\label{eq:MISE_bias}
\end{equation}

\noindent
and

\begin{equation}
<\frac{1} {N} \sum_{i=1}^{N}|{\bf F}_{true}({\bf x}_i)|^2>
\label{eq:MASE_bias}
\end{equation}

\noindent
respectively.
These quantities depend only on the mass distribution in the
configuration, and are independent of both the softening and the
number of particles, as borne out in Figure~\ref{repeat}.

In between the region dominated by the noise and that dominated by
the bias there is a minimum of the error. Since the noise decreases
considerably with $N$, while the bias is not affected by it, the
minimum error should decrease with increasing $N$ and should move to
smaller values of the softening, as is indeed shown to be the case in 
Figure~\ref{repeat}.

\subsection{The optimal value of the softening and the corresponding
value of the error}
\label{sec:plummer_opt}

Let us call
$\epsilon_{opt}$ the value of $\epsilon$ which gives the best
representation of the force, i.e. gives the lowest value of $MISE$ or
$MASE$ (which we will denote hereafter
by $MISE_{opt}$ or $MASE_{opt}$). Using
least square fits, M96 showed that power laws give good fits for the
values of $\epsilon_{opt}$ and $MISE_{opt}$ as a function of $N$, the
number of particles in the configuration. For $N$ between 30 and
300~000, and ${\cal C}$~=~1, A+98 obtained

$$\epsilon_{opt}=0.98N^{-0.26}$$

\noindent
and

$$MISE_{opt}=0.22N^{-0.68}$$

When few
particles are used (M96) the exponent of the 
$N$ dependence of $\epsilon_{opt}$ takes a value around -0.28 and
that of
$MISE_{opt}$ around -0.66. If we consider $N$ values between 10~000
and 300~000 then these values change to -0.23 and -0.76,
while the asymptotic values for
$N \rightarrow \infty$ are -0.2 and -0.8 (A+98). This argues that the
above equations 
are only approximate and that in fact the exponents are functions of
the number of particles considered. Nevertheless, within the range of 
particle numbers used in most collisionless $N$-body simulations they 
constitute a very good approximation.

\subsection{Comparing $MISE$ and $MASE$}
\label{sec:mise_mase}

$ISE$ and $ASE$ are, of course, only different ways of calculating the
same integral. In the case of radial $ISE$, however, the hypothesis is
implicitly made that 
the configuration is spherically symmetric, which is obviously true for 
the continuum distribution, or in the limit of an infinite number of
particles, but not, strictly speaking, in the case of a representation of
a Plummer sphere with a finite number of particles. A different way of
seeing the same effect is to say that many more points are used in the
sampling of the forces in case of an $ASE$ than in the case of this
$ISE$. Thus the $ASE$ ($MASE$) errors
are smaller than the corresponding radial $ISE$ ($MISE$) ones for the range
of values of 
$\epsilon$ considered e.g. in Figure~\ref{repeat}, except for the largest
values ($\epsilon$ greater than or of the order of 1.5), where the bias
predominates and the contribution from the noise is very small.
This advantage is linked to a corresponding disadvantage, namely that
the CPU time necessary for calculating an $ASE$ is larger than the 
corresponding time
for an $ISE$ calculation by a factor $(N-1)/n_{int}$, where $N$ the number of
particles in the representation and $n_{int}$ the number of points at
which the integrand is calculated. This makes the $MASE$ calculations
prohibitively expensive, even on powerful workstations. On the other
hand GRAPE-3 and GRAPE-4 are well adapted for such calculations. Thus
a $MASE$ calculation for 100~000 particles goes roughly 2000 times
faster on our GRAPE-3AF system than on its front end, an Ultra 2/200.

\subsection{Comparing results with GRAPE-3 and GRAPE-4}

Comparing the $MASE$ results calculated with GRAPE-3 and GRAPE-4,
respectively, we find that they agree very well.
This could, at first sight, seem at odds
with the fact that GRAPE-3 has a much more limited accuracy than
GRAPE-4 that
has near 64-bit precision. As argued, however, in A+98,  
the error in GRAPE-3 comes from round-off and can
therefore be considered as random. Thus when one adds the forces from
many particles 
this error cancels out and one obtains an accuracy similar to what is
obtained on GRAPE-4. For this
reason we have used both GRAPE systems 
for the calculations given in the next few sections.

\section{The case of two Plummer spheres }
\label{sec:twoplum}

\indent
In most simulations of galaxies or systems of galaxies it is necessary to
represent more than one component,  
each having considerably different properties. In order to assess the 
effect of this on the choice of the softening we will in this section 
consider the case of two non-truncated Plummer spheres. The first one has a
scale-length $a_1=1$ and the second one a scale-length $a_2=0.1$. We 
consider mass ratios of the two components $M_1/(M_1+M_2)=0$, 0.1, 
0.25, 0.5, 0.75, 0.9 and 1. The total mass is in all cases equal to
1, the total number of particles equal to 100~000 and the total number
of realisations equal to 40. We also consider two spatial  
configurations, one in which the two spheres are concentric and the 
other in which their centers are at a distance of 10 length units. For
some mass  
ratios the former configuration can be considered as representing
crudely a 
halo and bulge system, and the latter a halo of a target galaxy with 
a spherical companion. We will discuss
further only the case of the two concentric spheres, since  
the two configurations give essentially the same results for $MASE$.
As will be later shown in the next section it is the distance 
to the nearest neighbours that mainly determines $MASE$ and the 
corresponding $\epsilon_{opt}$, and this depends essentially on the
density of the Plummer sphere with the smallest scale-length, rather
than on its location. 

\begin{figure}
\includegraphics{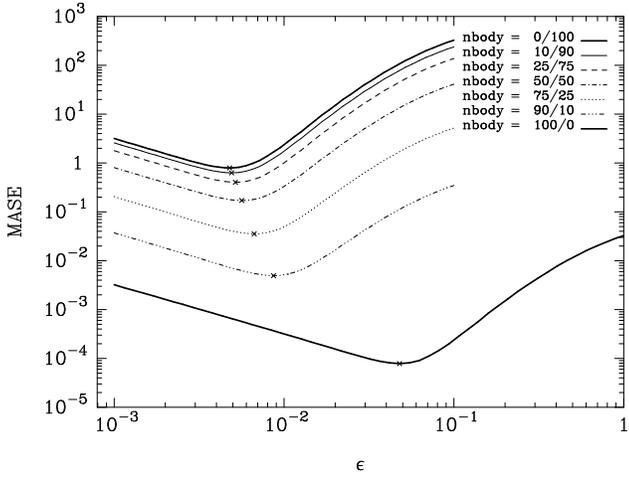} 
\vspace{7.5cm}
\caption{
$MASE$ as a function of $\epsilon$ for 100 000 particle
configurations. The lower heavy line corresponds to a Plummer sphere
of unit scale-length and the upper one to a Plummer sphere of 0.1
scale-length. The remaining ones correspond to cases where both Plummer
spheres are present, with mass ratios as given in the figure. The $\times$
symbols show the
positions of the minimum on each curve. As discussed in the text, for
these plots we have used ${\cal C}$~=~1,
since we are considering all components present in the same
simulation.
}
\label{twoplum_mase}
\end{figure}

\begin{figure}
\includegraphics{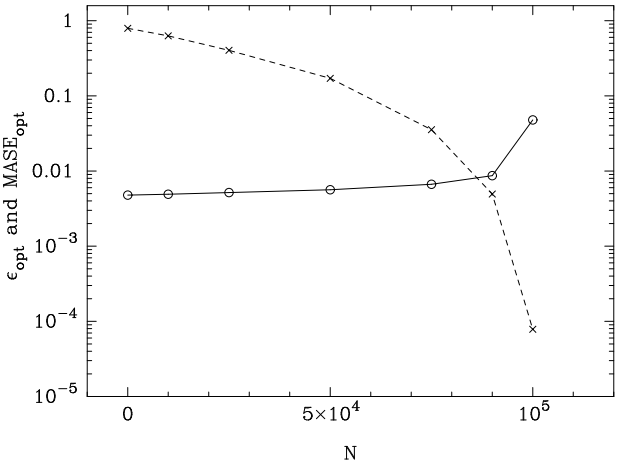} 
\vspace{7.5cm}
\caption{
Optimal softening (open circles) and corresponding $MASE_{opt}$ ($\times$) as
a function of the number of particles in the least dense
component. The points represented here are obtained from calculating
the minima of the curves of the previous figure.
}
\label{twoplum_eps1}
\end{figure}

Let us first discuss the case of a single simulation in which more
that one component is present. Since in this case we can not apply a
separate scaling to each component (cf. section~\ref{sec:notation})
we have to take ${\cal
C}$~=~1. Fig~\ref{twoplum_mase} compares the $MASE$ curves for the
seven mass ratios under consideration, and shows that more concentrated
configurations have 
larger errors and smaller corresponding $\epsilon_{opt}$ than less
concentrated ones. The
difference between the respective curves is considerable, since a
change of a factor of 10 in the scale-length makes a change of more
than $10^4$ in
$MASE_{opt}$ and of roughly 10 in $\epsilon_{opt}$. This figure also
shows that 
in an $N$-body simulation with many components the force on the densest
components will be less well represented than the force on the more
extended ones. The total error is bigger when the percentage of mass
in the denser component is larger. We also note that the 
increase in the error obtained by substituting 10\% of the particles in a 
loose configuration with a denser one is very big, whereas the 
decrease in the error obtained by substituting 10\% of the particles
in a dense  
configuration by a looser one is quite small. This is further stressed
in Figure~\ref{twoplum_eps1}, which is
obtained from the minima of the curves in Figure~\ref{twoplum_mase}, and
shows how the optimal softening 
depends on the percentage of mass in the least dense component. 
Similar calculations (not illustrated here) show that
there is hardly  
any difference between the cases where the two Plummer spheres are 
concentric and the case where they are separated, which leads to the
conclusion that  
the densest part influences the result always in the same way, 
independent of its location in the configuration.

\begin{figure}
\includegraphics{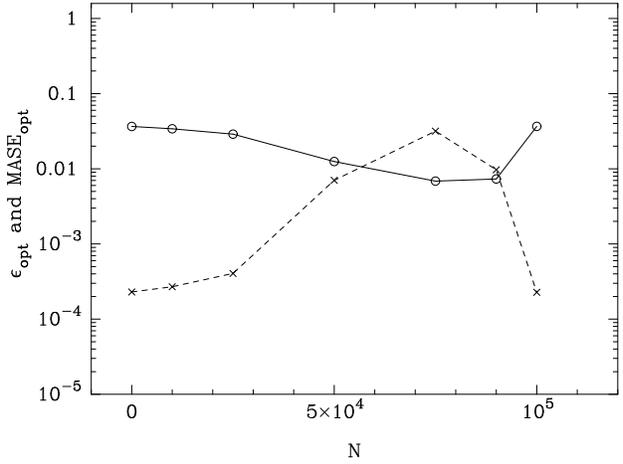} 
\vspace{7.5cm}
\caption{
Optimal softening (open circles) and corresponding value of the $MASE$
($\times$) as a function 
of the number of particles in the least dense component.
}
\label{twoplum_eps2}
\end{figure}

Now let us consider a different question and compare seven
different simulations, in each of which one of the above seven
configurations is present. Since now the scaling can be applied
independently to each of the configurations, in order to  compare
these cases between 
them we need to
use the weighted versions of the $MASE$ definition and
softening. Since the total 
mass of all the configurations is the same, the only factor that is
changing from one configuration to another is the half mass radius,
which takes respectively the values 0.13, 0.14, 0.18, 0.45, 0.97, 1.18
and 1.30.
Now the result of the least dense ($a_P=1$) and most dense ($a_P=0.1$)
configuration are of course identical. This simply reflects the 
fact that the densest Plummer sphere can be represented as well as
the least dense one, provided one uses appropriately weighted
softening values. This is clear also also from
Figure~\ref{twoplum_eps2} which shows $MASE_{opt}$ and
$\epsilon_{opt}$ as a function of the percentage of mass in the
least dense component. This figure also shows that, from the
configurations analysed here, the largest error corresponds to the
case with 25\% of the particles in the more concentrated Plummer
sphere. This is also the configuration for which $\epsilon_{opt}$ is
minimum. 

\section{The effect of central concentration}
\label{sec:homo_dehnen}

\subsection{Comparison of three different spherical distributions}
\indent

\begin{figure}
\includegraphics{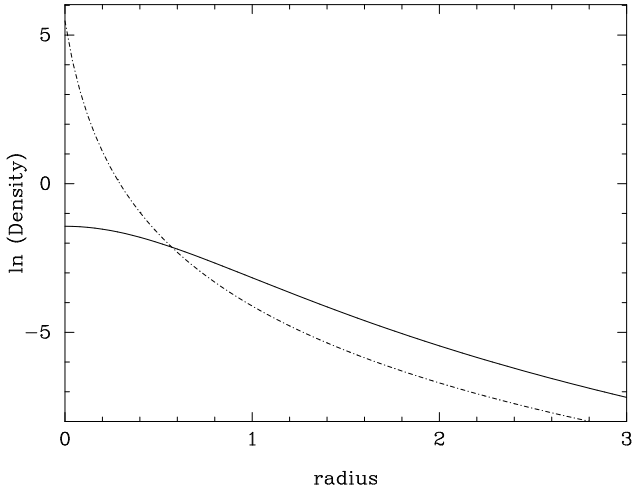}
\vspace{7.5cm}
\caption{
Comparing the density distributions of the Plummer sphere
(solid line) and the Dehnen $\gamma = 0$ sphere (dot-dashed line) used
in section \ref{sec:homo_dehnen}.  
}
\label{3dens}
\end{figure}

\begin{figure}
\includegraphics{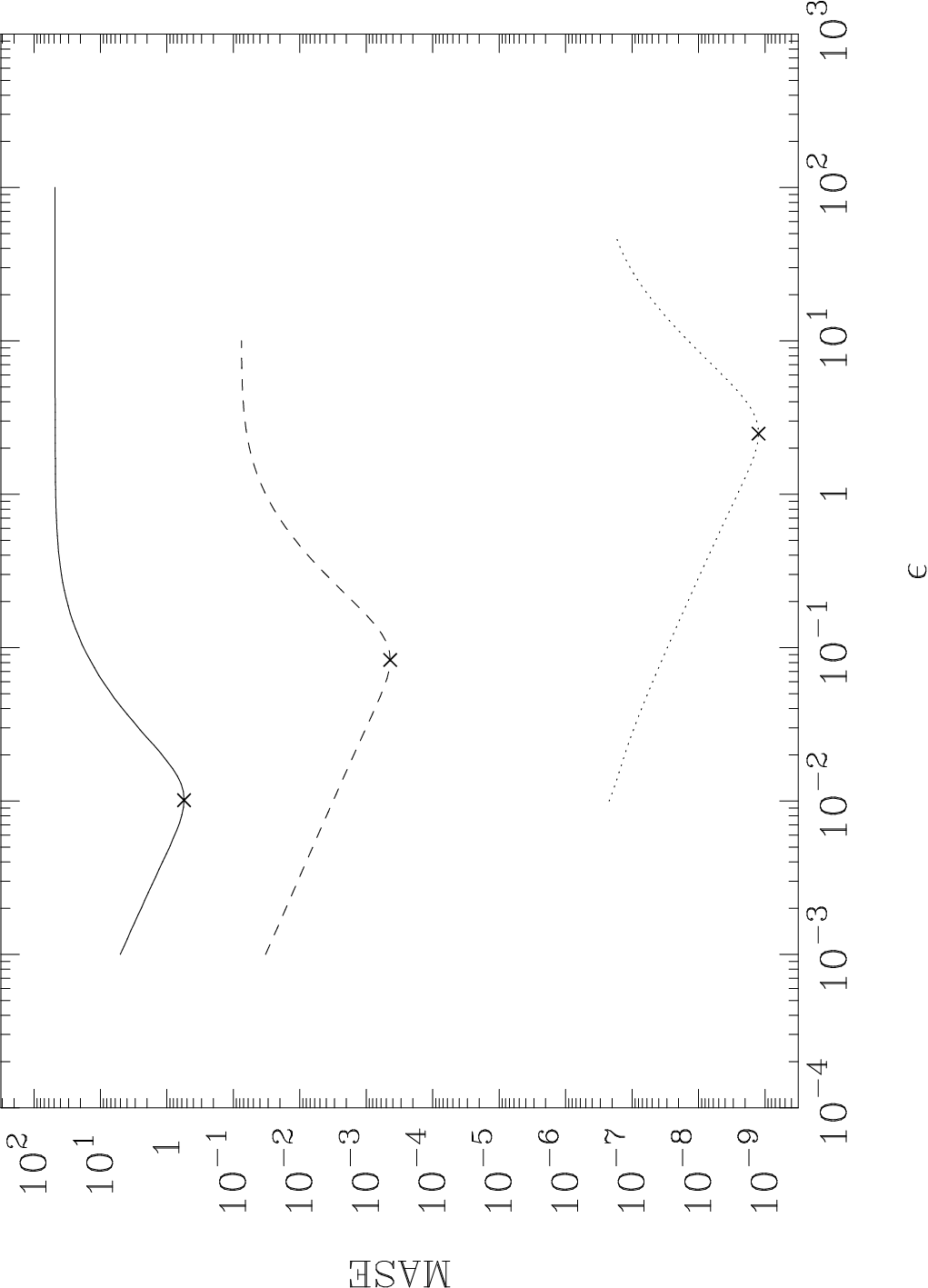}
\vspace{7.5cm}
\caption{
Comparing the unweighted $MASE$ as a function of $\epsilon$ for six
hundred 10~000 particle 
representations of a truncated homogeneous sphere (dotted line),
a Plummer sphere 
(dashed line) and a Dehnen sphere of index $\gamma = 0$ (solid line). 
}
\label{mase_hpd}
\end{figure}

\begin{figure}
\includegraphics{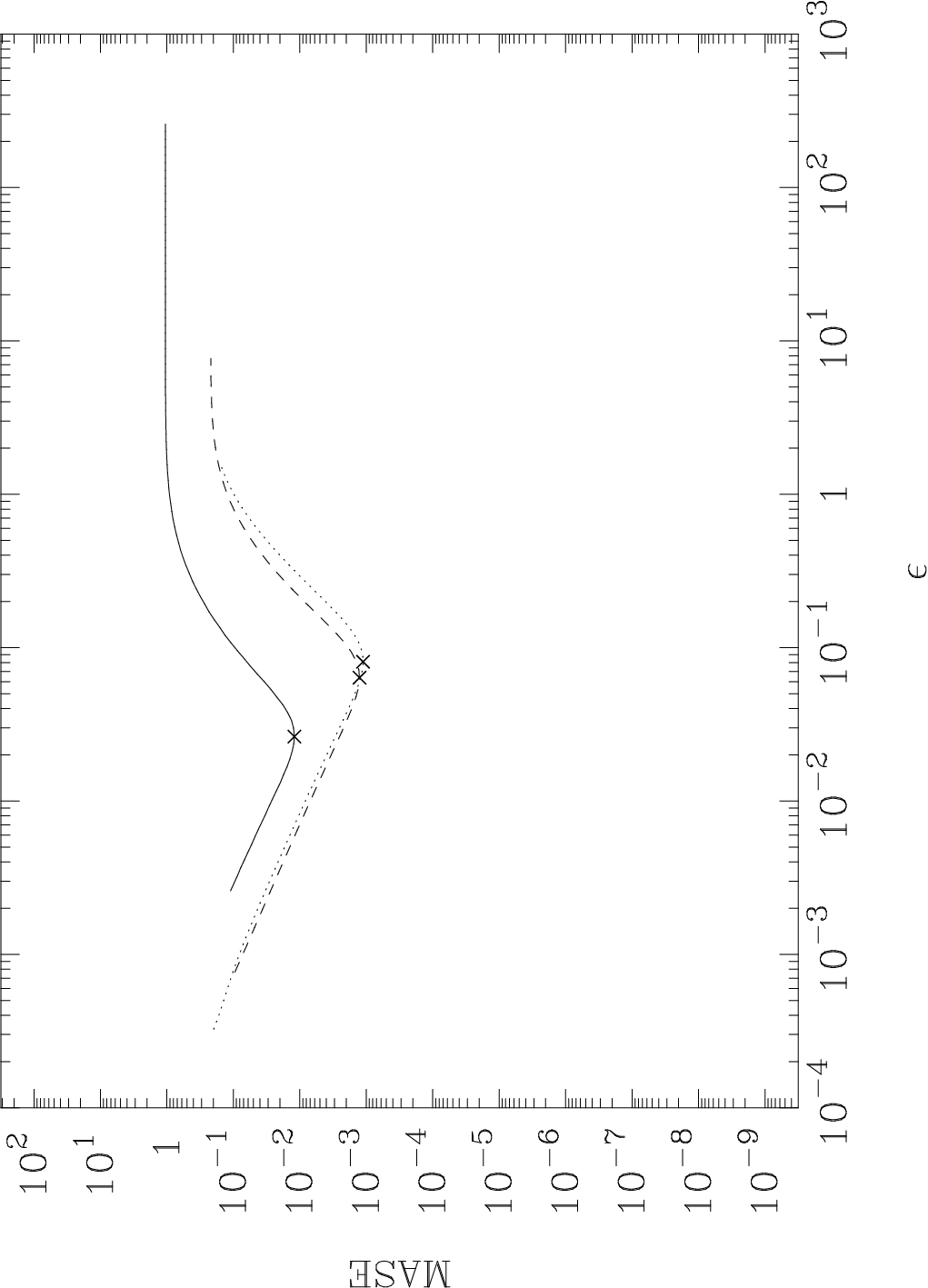}
\vspace{7.5cm}
\caption{
Comparing the weighted $MASE$ as a function of $\epsilon$ for six
hundred 10~000 particle 
representations of a truncated homogeneous sphere (dotted line),
a Plummer sphere 
(dashed line) and a Dehnen sphere of index $\gamma = 0$ (solid line). 
}
\label{mase_hpd_w}
\end{figure}

\begin{figure}
\includegraphics{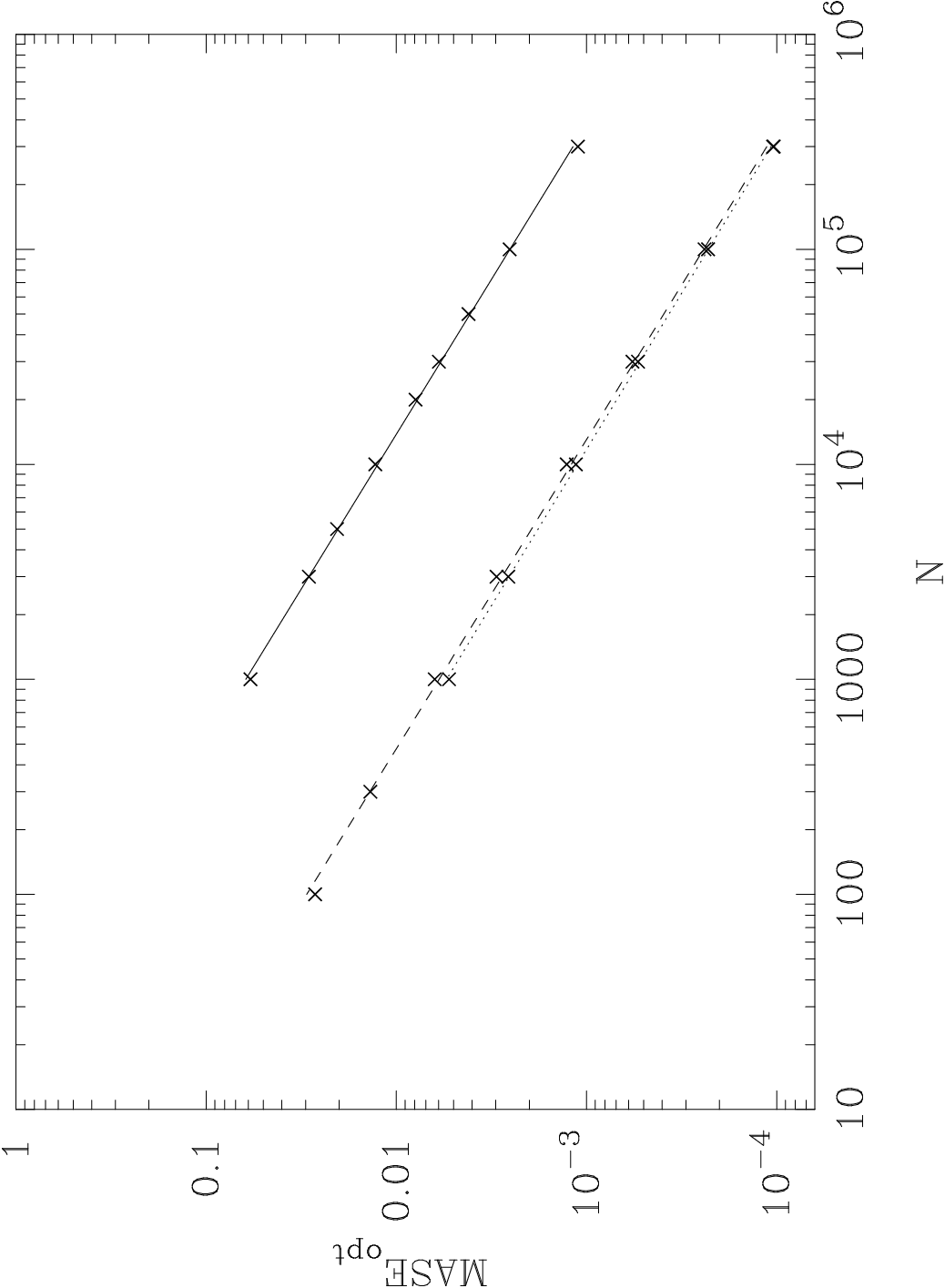}
\vspace{7.5cm}
\caption{
$MASE_{opt}$ as a function of $N$ for a truncated homogeneous sphere
(dotted line), a truncated Plummer sphere
(dashed line) and a truncated Dehnen sphere of index $\gamma = 0$
(solid line). 
}
\label{3MASEopt}
\end{figure}

\begin{figure}
\includegraphics{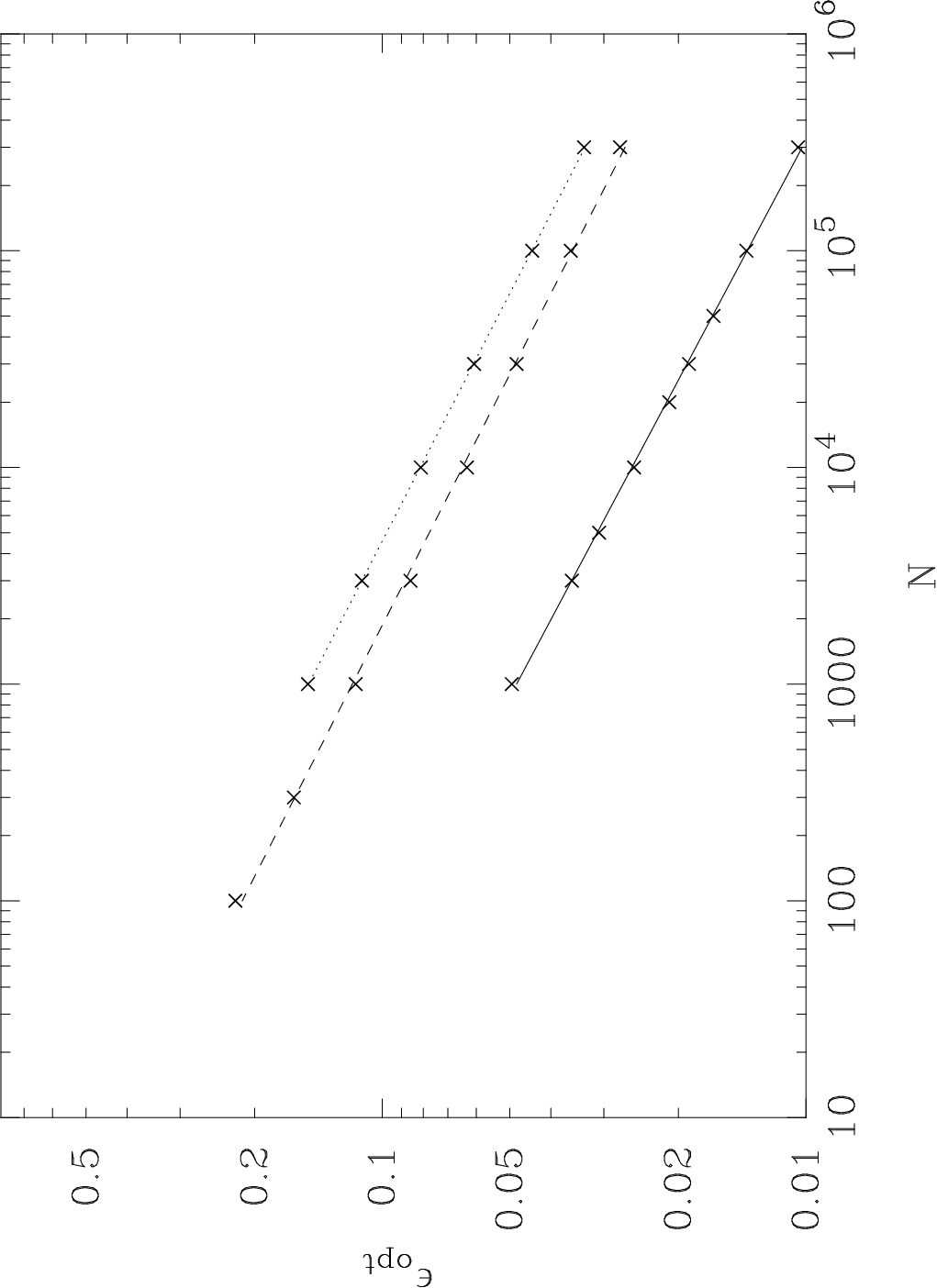}
\vspace{7.5cm}
\caption{
$\epsilon_{opt}$ as a function of $N$ for a truncated homogeneous sphere
(dotted line), a truncated Plummer sphere
(dashed line) and a truncated Dehnen sphere of index $\gamma = 0$ (solid line).
}
\label{3eopt}
\end{figure}

\begin{figure}
\includegraphics{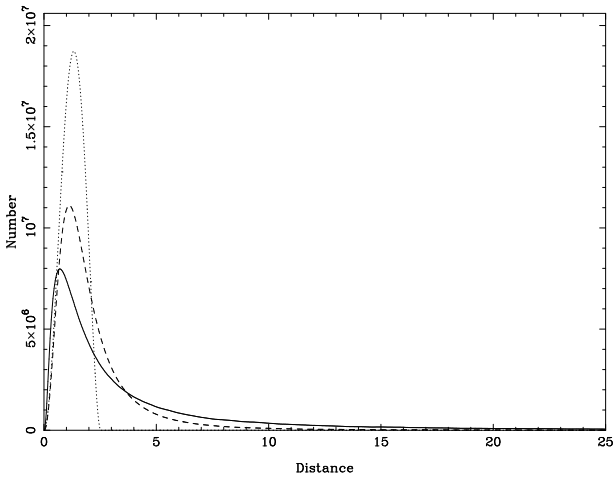}
\vspace{7.5cm}
\caption{
Histogram for inter-particle distances for the homogeneous sphere
(dotted line), the Plummer sphere (dashed line) and the
Dehnen $\gamma = 0$ sphere (solid line).
}
\label{histo_hpd}
\end{figure}

\begin{figure}
\includegraphics{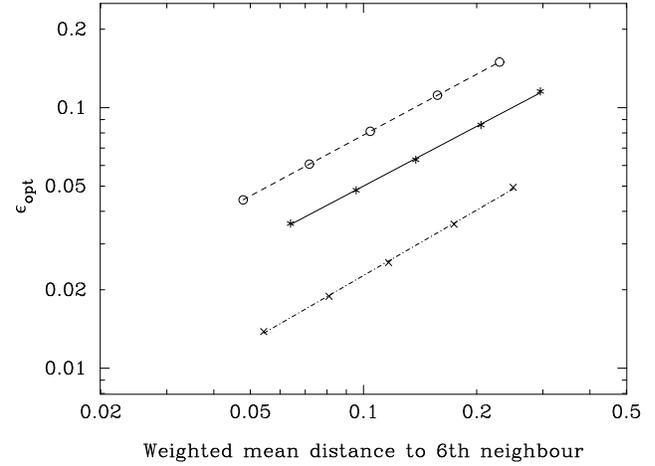}
\vspace{7.5cm}
\caption{
Optimal softening as a function of mean distance to the 6th nearest
neighbour. Values for the homogeneous sphere are given by open circles
and a dashed line, values for the Plummer sphere by asterisks and a full
line and values the Dehnen sphere by $\times$ symbols and a dot-dashed line.
}
\label{eopt_dist}
\end{figure}

We have so far considered the case of a Plummer sphere, a 
mass distribution frequently used in astrophysics. In this section we
will consider two other density distributions, a truncated homogeneous
sphere and a Dehnen sphere (Dehnen 1993). The former is less centrally
concentrated than the Plummer sphere and the latter more, so that
by comparing the respective $MISE$ or $MASE$ we can test the effect of
central concentration on the optimal $\epsilon$ and on the corresponding 
accuracy. Since both Plummer and Dehnen spheres extend to infinity, we
have introduced in both cases a cut-off radius, taken so that the mass
within that radius is equal to 0.999 times the total mass. 

The density profile of the homogeneous sphere is 

\[\rho (r) =\left\{ \begin{array}{ll}
3 M_T/{4 \pi R^3} & \mbox{$r \leq R$} \\
0 & \mbox{$r > R$} \\
\end{array}
\right. \]

\noindent
where $R$ is its outer radius and $M_T$ its total mass. For the Dehnen
sphere we have

\[\rho (r) =\left\{ \begin{array}{ll}
{{(3-\gamma) M_T} \over {4 \pi}} {{a_D} \over
{r^{\gamma}(r+a_D)^{(4-\gamma)}}} & \mbox{$r \leq R$} \\
0 & \mbox{$r > R$} \\
\end{array}
\right. \]

\noindent
where $M_T$ is its total mass, $a_D$ is its scale-length
and $\gamma$ is its concentration index, determining the slope of the
density at the origin. 
For the examples discussed below we took $a_D$~=~0.1 and $\gamma$~=~0.

The truncation radii, containing 0.999 of the total mass, are
equal to 299.8 and 38.71 for the Dehnen and Plummer configuration
respectively. For the homogeneous sphere we have also taken $R$~=
38.71 and in all three models we have taken the mass within the
truncation radius to be equal to 1. The Plummer and Dehnen density
distributions are compared in Figure~\ref{3dens}. 

Figure~\ref{mase_hpd} and \ref{mase_hpd_w} compare the results for $MASE$ 
obtained from six hundred representations of 10~000 particles each. In
the former we use the un-weighted definition of $MASE$ (i.e. with ${\cal
C}$ = 1) and of the softening,
and in the latter the weighted definition. We note that, in both
cases, the Dehnen sphere, which is the most concentrated of the 
three configurations, requires smaller softening values and is always
less accurately represented than the two more
spread out configurations, in good agreement with the results of section
\ref{sec:twoplum}. 
The differences, however, are much more important
in the comparison of the un-weighted $MASE$. Furthermore in this case the
differences between the Plummer and the homogeneous sphere are very
large, while when the comparison is between the weighted functions the
difference is very small. Again this is in agreement with the results
of section \ref{sec:twoplum}. 
If we simulate all three spheres in the same configuration we have to 
use for all cases the same softening and ${\cal C}$ = 1. 
Now the forces of the Dehnen sphere will be very badly
represented and those of the homogeneous sphere very well.
The situation is totally different if we are 
simulating one of those spheres only, because then we have to
calibrate
the lengths appropriately, so that all systems have the same half-mass
radius.
Now the Dehnen sphere will do
much better than in the un-weighted case, and the homogeneous sphere
much worse. Thus the weighted $MASE_{opt}$
is respectively 
0.001, 0.001 and 0.01 for the homogeneous, Plummer and Dehnen  spheres.
In particular the differences between the Plummer and the
homogeneous sphere are very small.
The $MASE_{opt}$ for the Dehnen sphere is considerably larger than for
the other 
two, presumably because, in the Dehnen sphere, we are trying with a
single value of the 
softening to accommodate both very dense and very sparse regions.
 
Figures~\ref{3MASEopt} and \ref{3eopt} compare weighted $MASE_{opt}$ and
$\epsilon_{opt}$ as a function of $N$ for the three configurations and
confirm the trend seen in Figure~\ref{mase_hpd_w}. The dependence of
$MASE_{opt}$ and 
$\epsilon_{opt}$ on $N$ for all three density distributions can be
represented by power laws  

\begin{equation}
MASE_{opt}=BN^b
\label{eq:mase_3d}
\end{equation}

\noindent
and

\begin{equation}
\epsilon_{opt}=AN^a
\label{eq:eopt_3d}
\end{equation}

The values of the coefficients are given in Table~\ref{tab:hpd}. 
Since the exponent in the above dependences depends somewhat on the
range of $N$ used (cf. A+98 and section \ref{sec:plummer_opt}) , we
used for the three mass distributions the same values of $N$ for the
linear fits, namely the values for $N$ = 1~000, 3~000, 10~000, 30~000,
100~000 and 300~000. 

\begin{table}
\centering
\caption{Coefficients of the power laws in equations \ref{eq:mase_3d}
and \ref{eq:eopt_3d} }
\label{tab:hpd}
\begin{tabular}{@{}lcccc@{}}
Mass distribution & A & a & B & b \\ 
Homogeneous sphere & 28. & -0.26 & 7.1 $\times 10^{-7}$ & -0.69 \\
~~~~~(un-weighted) &&&&\\
Plummer sphere & 0.84 & -0.25 & 0.32 & -0.72 \\
~~~~~(un-weighted) &&&&\\
Dehnen $\gamma$ = 0 sphere & 0.12 & -0.27 & 340. & -0.69 \\
~~~~~(un-weighted) &&&&\\
Homogeneous sphere & 0.92 & -0.26 & 0.63 & -0.69 \\
~~~~~(weighted) &&&&\\
Plummer sphere & 0.64 & -0.25 & 0.94 & -0.72	\\
~~~~~(weighted) &&&&\\
Dehnen $\gamma$ = 0 sphere & 0.32 & -0.27 &  7.5 & -0.69 \\
~~~~~(weighted) &&&&\\
\end{tabular}
\end{table}

The table and figures show clearly that more centrally concentrated
configurations need smaller 
values of the softening for an optimal representation of the force, and
the precision achieved is never as good as for less centrally
concentrated configurations. The minimum error, $MASE_{opt}$,
decreases somewhat faster with $N$ for the case of the Plummer sphere,
but the differences are 
small.

Figure~\ref{histo_hpd} compares the
histograms of the inter-particle distances for the three models after
they have been rescaled so that the half-mass radii are in all three
cases the same. This scaling is appropriate for understanding the
results obtained with weights. Each
histogram has been obtained from ten 10~000 
particle realisations of each model. We note that the peak of the
histogram is nearest to the center for the Dehnen sphere, followed by
the Plummer sphere, while the peak of the histogram for the homogeneous
sphere is yet further out. It is thus expected that there are more
particles very close to each other in the Dehnen sphere than in the
Plummer one, and even more compared to the homogeneous one. However
the Dehnen sphere has also more particles with very large
inter-particle distances, while these are fewer for the Plummer
sphere and even more so for the homogeneous sphere.

\subsection{Extension to other configurations}
\indent

The results shown in Figure~\ref{3eopt} give us
the optimal value of the softening, which 
gives the best representation of the forces, as a function of
the number
of particles $N$. However they also show that a value of softening
which is optimal for one type of mass distribution is not necessarily
optimal for 
another, and that the optimal value depends strongly on the central
concentration of the distribution. Thus, in order to find
$\epsilon_{opt}$ for a mass distribution
other than the three discussed here, one can either do the full 
calculations of the $MASE$, as above, or use the above results to
obtain rough estimates. Since the
former is rather demanding, we would like, in the remaining of this
section, to see whether it is possible to obtain some, albeit
crude, estimates of the 
optimal softening as a function of quantities linked to the mass
distribution, but more straightforward to calculate than the $MASE$.

As we have already seen, smaller values of the softening are necessary
for more compact configurations or for representations with a larger
number of particles, i.e. in cases where the particles are
closer together. This suggests that some measure of inter-particle
distances could be used for estimating the optimal value of the
softening. Furthermore, since $MASE$ is more dependent on the accuracy
of the forces to nearby particles, we will try using the 
distances of the few nearest neighbours. Let us thus, for a given
configuration, measure, for every particle, the distance to its twelve
nearest neighbours. Now we need to average this over all particles in
the configuration, in order to obtain, for the whole configuration, 
the mean distance to the nearest neighbour, the mean distance to the
second nearest neighbour etc., up to the mean distance to the 12th
nearest neighbour. A standard arithmetic average would not be appropriate for
this. This can be understood if we mentally add to the configuration a
single particle, 
located so far from it that it can, for argument's sake, be considered
at infinity. This new particle will not influence the value of $MASE$,
the value of $\epsilon_{opt}$ that should be used, or the accuracy in
the force calculations. On the other hand it will influence the mean
distance to the $k$th neighbour. It is thus not reasonable to expect a
close relation between $\epsilon_{opt}$ and the arithmetic mean of the
$k$th closest neighbour of all particles. For this reason, instead of the
standard arithmetic mean, we will prefer using
the harmonic mean

$$ r_{k,mean1} = (N^{-1} \sum_{i=1}^{N} r^{-1}_{k,i})^{-1}$$

\noindent
or, since the force is inversely proportional to the square of the
distance, 

$$ r_{k,mean2} = (N^{-1} \sum_{i=1}^{N} r^{-2}_{k,i})^{-1/2}$$

\noindent
where the summations are over all the particles in the configuration and
$k = (1,..., 12)$ .
In this way we obtain for a given configuration 
the mean distance to the nearest neighbour, the mean distance to the
second nearest neighbour etc., up to the mean distance to the 12th
nearest neighbour. It is possible to diminish the noise in such
calculations by considering several realisations of the same
configuration and then mean $r_{k,mean1}$ or $r_{k,mean2}$, now using
simple arithmetic means over all realisations. We calculated these mean
inter-particle distances as a function of $N$ for the three density
distributions considered above, using 3 $\times 10^6 /
N$ realisations. In all cases the dependence is roughly linear on a
log-log plane, and thus can be represented by power laws
of the type

$$r_{k,mean}=AN^a$$

\noindent
where $r_{k,mean}$ is equal to $r_{k,mean1}$ or $r_{k,mean2}$. In both
cases the value of $a$ is around -0.33 or -0.34 and does not depend
much on which of the twelve nearest neighbours is considered. The
latter is not true if we use standard means. The values of A
depend of course on which of the nearest neighbours is chosen. 

We have thus obtained so far, for a given density distribution and a given
number of particles $N$, the average distance to the $k$th nearest
neighbour, for $k$ between 1 and 12, as well as an optimal softening
$\epsilon_{opt}$ 
(cf. Figure~\ref{3eopt}). From these two we can eliminate the
dependence on $N$ and obtain the dependence of $\epsilon_{opt}$
directly on the average distance to the $k$th nearest neighbour. This
is given in Figure~\ref{eopt_dist} for the sixth neighbour, after
both distances and the 
softening have been weighted appropriately, as discussed in section
\ref{sec:notation}. We have repeated this
exercise for all other values of $k$ between 1 and 12, but since the
results are similar we do not reproduce them here. Again a power law
gives a good representation of the dependences 

\begin{equation}
\epsilon_{opt}=Ar_{k,mean1}^a
\label{eq:eopt_est}
\end{equation}

\noindent
The values of A and a, for various neighbours are given in Table
\ref{tab:neib_hpd}. 

\begin{table}
\centering
\caption{ Coefficients of the power laws in equation (\ref{eq:eopt_est}) }
\label{tab:neib_hpd}
\begin{tabular}{@{}ccccccc@{}}
 & \multicolumn{2}{|c|}{Homogeneous} & \multicolumn{2}{|c|}{Plummer} & \multicolumn{2}{|c|}{Dehnen} \\
Neighbour & A & a & A & a & A & a \\ 
  1  & 0.95 & 0.78  & 0.55 & 0.76 & 0.31 & 0.83 \\
  3  & 0.59 & 0.78  & 0.35 & 0.76 & 0.19 & 0.83 \\
  5  & 0.50 & 0.78  & 0.30 & 0.76 & 0.16 & 0.83 \\
  7  & 0.45 & 0.78  & 0.28 & 0.76 & 0.15 & 0.83 \\
  9  & 0.41 & 0.77  & 0.26 & 0.76 & 0.14 & 0.84 \\
  11 & 0.39 & 0.77  & 0.25 & 0.76 & 0.13 & 0.84\\
\end{tabular}
\end{table}

Figure~\ref{eopt_dist} can give some, albeit rough, estimate of the
optimal softening to be used, once an $r_{k,mean}$ has been
calculated, since the differences between the three models are less,
or of the order of, 0.5 dex. It is, however, possible to narrow down
the prediction further. Indeed the three dependences are ordered as a
function of the central concentration of the corresponding models,
less concentrated models corresponding to higher softening than more
concentrated ones, in good agreement with what was previously
discussed in this section. Thus comparing the central concentration of
the new model, whose optimal softening we want to estimate, with that
of the three considered here, should probably narrow the estimate to
0.2 or 0.3 dex.

\section{The effect of triaxiality }
\label{sec:ferrers}
\indent

\begin{figure}
\includegraphics{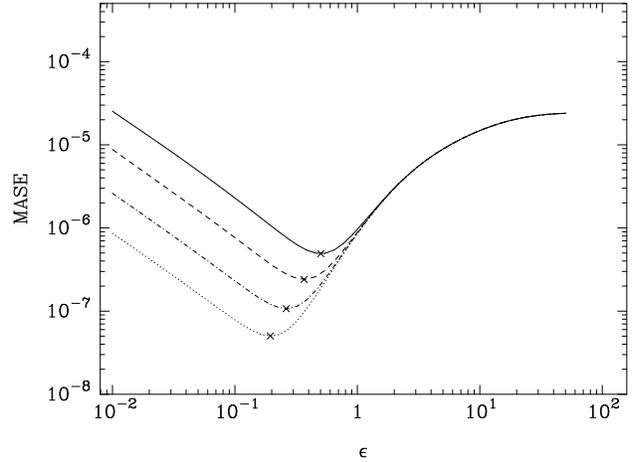} 
\vspace{7.5cm}
\caption{
$MASE$ as a function of softening for a homogeneous Ferrers
ellipsoid of axial ratio 10:10:1. The various curves correspond to a
different number $N$ of particles in the configuration; 10~000 (solid
line), 30~000 (dashed line), 100~000 (dot-dashed line), and 300~000
(dotted line), respectively. 
}
\label{ferrers_N}
\end{figure}

\begin{figure}
\includegraphics{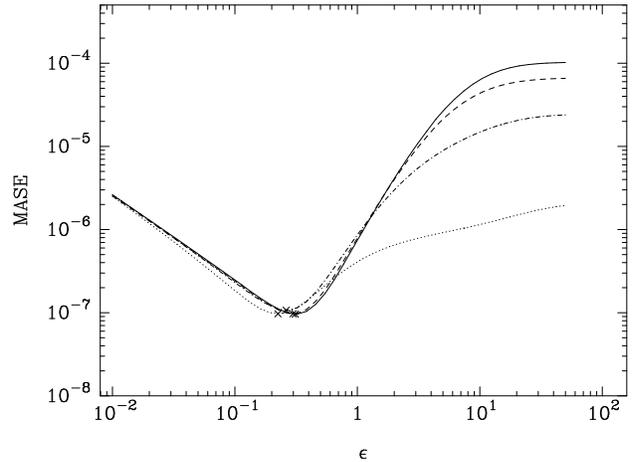} 
\vspace{7.5cm}
\caption{
$MASE$ as a function of softening for four Ferrers'
ellipsoids of axial ratio 1:1:1 (solid line), 12:4:3 (dashed line),
10:10:1 (dash-dotted line) and 74:74:1 (dotted line). In all cases the
number of particles is equal to 100~000 and the number of realisations
is equal to 60.
}
\label{ferrers_shape}
\end{figure}

\begin{figure}
\includegraphics{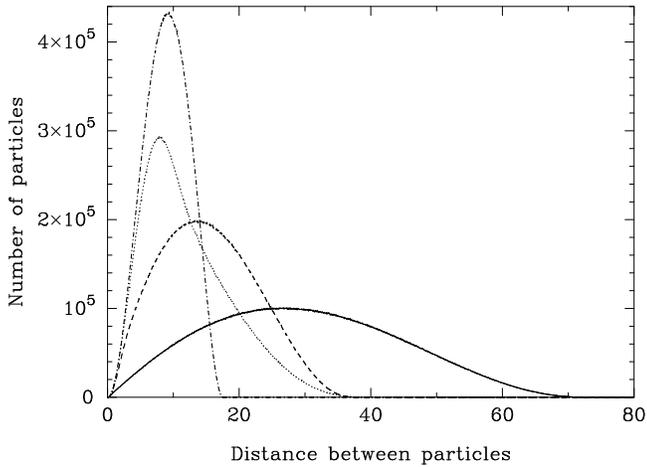} 
\vspace{7.5cm}
\caption{
Histogram of inter-particle distances of all pairs of
particles in Ferrers' homogeneous ellipsoids with axial ratio
1:1:1 (dot-dashed line), 12:4:3 (dotted line), 
10:10:1 (dashed line) and 74:74:1 (solid line). Each
histogram was obtained from ten 10~000 particle realisations. 
}
\label{ferrers_histo}
\end{figure}

\begin{figure}
\includegraphics{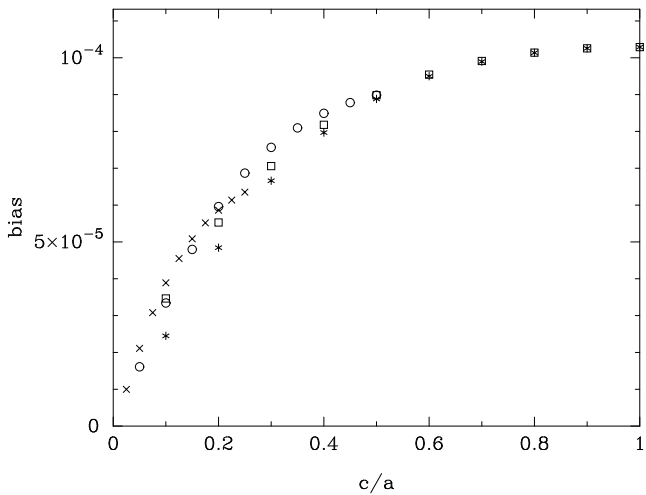} 
\vspace{7.5cm}
\caption{
Bias for a softening tending to infinity as a function of the axial
ratio $c/a$ of the ellipsoid. Asterisks correspond to oblate
ellipsoids and open squares to prolate ones. Open circles and $\times$
correspond to axial ratios $a:a/2:c$ and $a:a/4:c$ respectively.
All ellipsoids have the same volume as the four studied in more detail
in this section.
}
\label{ferrers_bias}
\end{figure}

So far we have considered only spherical objects. By using,
however, the 
$MASE$ rather than the radial $MISE$ accuracy estimator it is possible to
consider non-spherical configurations. As an example we will in this
section consider a Ferrers' ellipsoid (Ferrers 1877), 
a distribution
often used for modeling bars. The corresponding density is

\[\rho=\left\{ \begin{array}{ll}
\rho_{0} (1-g^2)^n & \mbox{if $g<1$} \\
0 & \mbox{otherwise}
\end{array}
\right. \]

\noindent
where $g^2=x^2/a^2+y^2/b^2+z^2/c^2$, $a>b>c$ are the major,
intermediate and minor axes, $\rho_0$ is the central
density of the ellipsoid and $n$ is an index determining the mass
distribution in the ellipsoid. We calculated $MASE$ values for four such
ellipsoids with $n=0$ and 
$a:b:c$ = 1:1:1, 12:4:3, 10:10:1 and 74:74:1 respectively. These
represent objects common in an astrophysical context: a
sphere, a bar, a thin and a very thin disc. 
Undoubtedly the disc component in
real galaxies is not as thin as the thinnest of the two discs that we
are using here, but we have added  
on purpose this rather extreme ellipsoid to see how an extreme thinness would
affect the values of $MASE$. To facilitate comparisons, the
four ellipsoids have been taken to have the same volume and mass
(equal to 666.666 and 1 respectively), so we can take ${\cal C}$~=~1.   

Figure~\ref{ferrers_N} shows $MASE$ as a function of the softening
for a Ferrers homogeneous ellipsoid of axial ratio 10:10:1 and
different values of $N$. The general outline of the curves is the same as for 
the spherical objects (cf. section \ref{sec:plummer_MASE}). As is the case
for the spherical distributions $\epsilon_{opt}$ can be represented by
a power law dependence on the number of particles $N$. 

In order to evaluate better the effect of triaxiality on the error in
the force calculation we plot in Figure~\ref{ferrers_shape} the
$MASE$ as a function of $\epsilon$ for the four ellipsoids under
consideration. In order not to burden the figure we plot only one
value of $N$, in this case 100~000. Similar results have been obtained
for the other values of $N$ considered. Let us first concider
the flat part of the
curve, where the $MASE$ is practically independent of the softening,
and which occurs for large values of $\epsilon$. For brievity we will
hereafter call this the bias
part of the curve. Two effects in particular are clear from
Figure~\ref{ferrers_shape} about the bias part. The first concerns for how
big a value of the softening this part is attained and the second what the 
value of $MASE$ on this section is. 

Figure~\ref{ferrers_shape} shows that the bias part occurs for relatively
smaller $\epsilon$ values when the shape is spherical, and
considerably larger ones when the departure from sphericity is
large. This can be understood with the help of
Figure~\ref{ferrers_histo}, which compares histograms of the
inter-particle distances for the four ellipsoids under
consideration. We note that, as expected, the higher the departure
from sphericity, the larger the percentage of large inter-particle
distances. As discussed in section \ref{sec:plummer_MASE}, the bias
dominates when the softening is larger than most inter-particle
distances. According to Figure~\ref{ferrers_histo} this
is expected to happen for larger softenings for objects that depart
more from sphericity and this is indeed verified in
Figure~\ref{ferrers_shape}.

The second effect that is clear from Figure~\ref{ferrers_shape} is
that the value of the bias, or, in other words, the values of $MASE$
on the bias part of the curve,
also depends on the departure from sphericity. In particular the larger
the departure from sphericity, the smaller this value is. This can
again be understood with the discussion in section
\ref{sec:plummer_MASE}. We can simply calculate the value of
the bias at 
infinite softening from equations (\ref{eq:MISE_bias}) or
(\ref{eq:MASE_bias}) by integrating the force of a homogeneous prolate
spheroid appropriately over its volume. We find the highest biases for
the most spherical objects, and the values decrease as the departure
from sphericity increases, in good agreement with the results of 
Figure~\ref{ferrers_shape}. Some examples are shown in
Figure~\ref{ferrers_bias}. We see that both oblate and prolate objects
have high values of the bias when they are nearly spherical. When they
are far from spherical then the bias for the prolate cases is somewhat
higher than the bias for the corresponding oblate ones. 
Figures~\ref{ferrers_shape} and \ref{ferrers_bias} show that the effect of
shape on the bias can be quite important, changing the corresponding
values of $MASE$ by over an order of magnitude. 

Now let us turn to the values of $MASE$ corresponding to smaller
softenings. Figure~\ref{ferrers_shape} shows that they do not depend
much on the shape. This
can be understood since, for such values of the softening, it is the noise that
affects the error most, and this should not depend on the shape of the
object. In particular for values around $\epsilon_{opt}$ this
dependance is negligible. Thus we can conclude that the shape of the
homogeneous Ferrers  
ellipsoid does not influence much either the value of $\epsilon_{opt}$ which
brings the best representation of the forces, or the corresponding
value of the error $MASE_{opt}$. 

\section{Different forms of softening }
\label{sec:other-soft}
\indent

\begin{figure}
\includegraphics{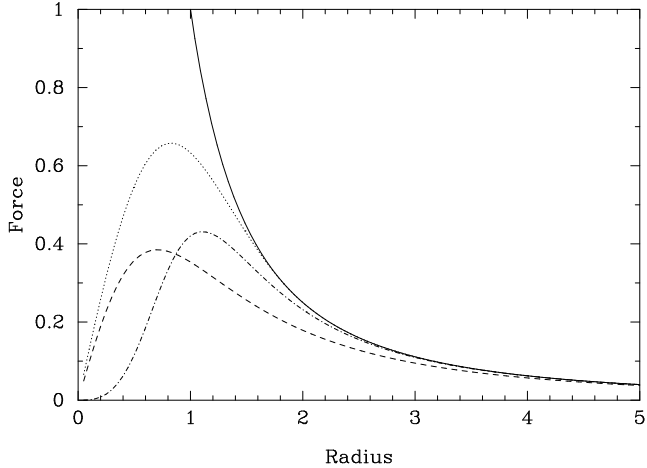}
\vspace{7.5cm}
\caption{
Comparison of the forces calculated with three different types of
softening to the Newtonian force (solid line). The Plummer
softening is given by a dashed line, the power softening with $p$=4
with a dot-dashed one, and the spline softening with a dotted line.
}
\label{compare_soft}
\end{figure}

\begin{figure}
\includegraphics{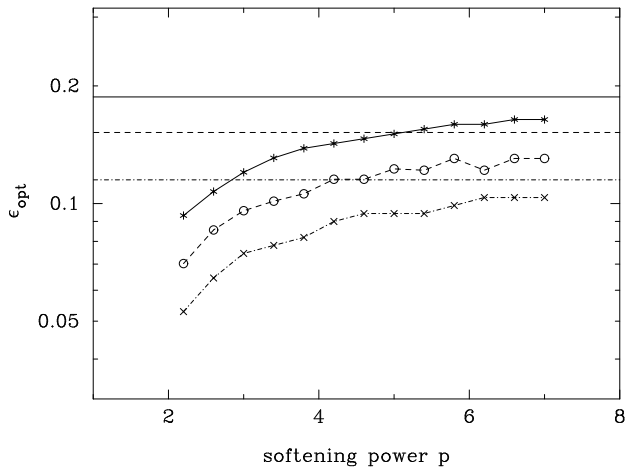}
\vspace{7.5cm}
\caption{
Optimal softening  
as a function of the exponent $p$ in the power law softening. 
The solid line and stars correspond to $N$ = 10~000, the dashed
line and open circles correspond $N$ = 30~000, and the dot-dashed line
and $\times$ symbols to $N$ = 100 000 (cf. text).
The corresponding values for the 
spline softening are given by horizontal lines of the same type.
}
\label{eps_p_spline}
\end{figure}

\begin{figure}
\includegraphics{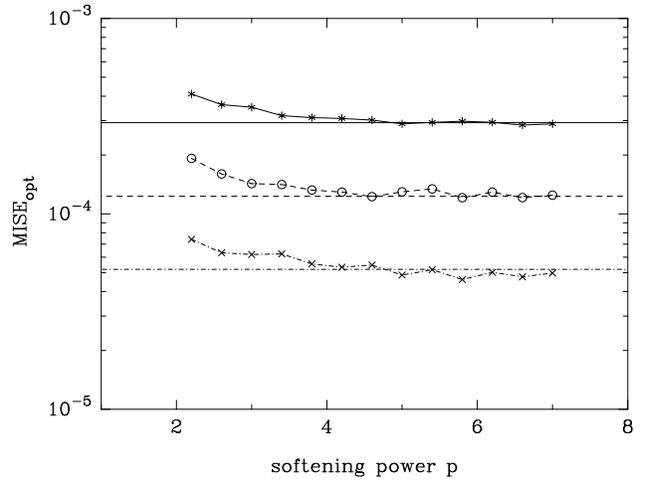}
\vspace{7.5cm}
\caption{
As in the previous figure, but for the minimum
value of the radial $MISE$.
}
\label{MISE_p_spline}
\end{figure}

\begin{figure}
\includegraphics{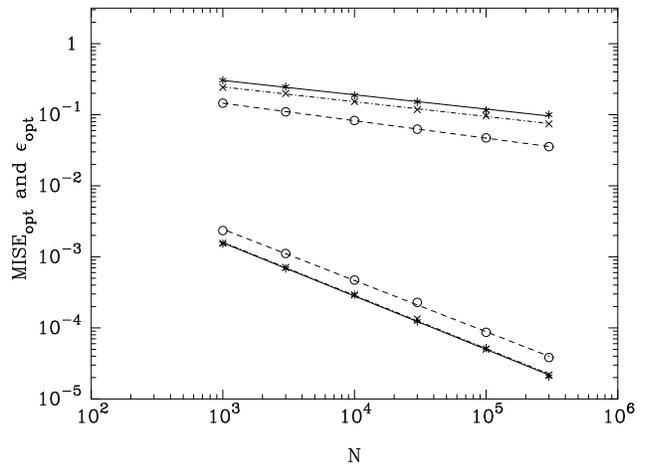}
\vspace{7.5cm}
\caption{
Optimal softening and corresponding minimum value of the radial $MISE$
as a function of $N$ for a Plummer sphere. This figure compares the 
results for a spline softening (solid line), a power softening for $p$
= 2 (dashed line), and a power softening for $p$ = 5 (dot-dashed line). 
}
\label{N_p_spline}
\end{figure}

In all calculations presented so far we have used the standard Plummer
softening, introduced in section \ref{sec:notation}.   
There are, however, many alternative ways of 
introducing the smoothing. For example, instead of using the second
power of the softening and of the inter-particle distance in equation
(\ref{eq:force_plum}),
one could use any other value, including non-integer.   
Another alternative is to use, instead of the force in a Plummer
sphere, the force in a sphere of constant density (e.g. Pfenniger \&
Friedli 1993, Palou\v{s}, Jungwiert \& Kopeck\'y 1993), or of any other
radial dependence (e.g. Palou\v{s}, Jungwiert \& Kopeck\'y
1993). Alternatively one could substitute the force between two point
masses for the force between two spheres. A few density profiles of
the mass within the sphere lend
themselves to an easy calculation of the potential and corresponding
force, in particular certain radial profiles given by
polynomials. Such examples have been discussed by Hockney \& Eastwood
(1981) and by Dyer \& Ip (1993). 

We will discuss in this section some alternative types of softening
and compare how well 
they fare in the $MISE$ test. Since both GRAPE-3 and GRAPE-4
calculate the force only with a Plummer softening, we performed all
calculations in this section on workstations, using radial $MISE$, rather than
$MASE$, and 64 bit precision. The integration in radial $MISE$ was done as
described in section~\ref{sec:notation}.

As a first example of an alternative force calculation we
will use the spline approximations given by Hernquist \& Katz (1989),
namely ${\bf F}=-m{\bf r}f(r)$, where

\vspace{5mm}
\noindent
%\begin*{equation}
\begin{minipage}{16cm}

\tiny
\[f(r)=\left\{ \begin{array}{lll}
{1/{\epsilon}^3}[(4/3)-(6/5)u^2+(1/2)u^3]  
         & \mbox{$0 \leq u \leq 1$} \\
1/r^3 [-1/15+(8/3)u^3-3u^4+(6/5)u^5-(1/6)u^6] 
	& \mbox{$ 1 \leq u \leq 2$} \\
1/r^3 & \mbox{$u \geq 2$}
\end{array}
\right. \]

\normalsize
\end{minipage}
%\end*{equation}

\vspace{5mm}
\noindent
and $u=r/\epsilon$.

Note that for $r \geq 2\epsilon$ the value of the force in this
approximation is exactly the Newtonian force. In other
words contrary to the Plummer softening, and to the power-law
softening introduced below, the spline softening is approximate only
for small distances.
  
As a second example we will consider an extension of the Plummer
softening to values of the exponent other than two. This can be given
by 

\begin{equation}
{\bf F}_i = G \sum_{j=1}^{N} 
\frac {m_j({\bf x}_j - {\bf x}_i) |{\bf x}_i - {\bf x}_j|^{p-2}} 
{(|{\bf x}_i - {\bf x}_j|^p + \epsilon ^p)^{1/p+1}}
\label{eq:force_gplum}
\end{equation}

\noindent
which for $p$=2 gives back equation (\ref{eq:force_plum}).
A point to note is that for all values of $p$, including the commonly
used value $p$ = 2, these forces tend to the
Newtonian one only asymptotically, i.e. even at large distances there
is a finite, albeit small, difference between the results they give and
those of the Newtonian force. In other words they introduce a small but
non-zero smoothing even at large distances, where it isn't necessary.

Figure~\ref{compare_soft} compares the amplitude of the non-softened
Newtonian force 
with those of the Plummer, the $p$ = 4 power-law and spline softened
forces. For this figure we have taken  
the softening as well as the masses equal to unity.
We note that the force that
approximates best the Newtonian one is the spline, followed by the
higher power softening, while the Plummer softening does the least
well of the three. Thus the forces agree better than 5\% with the
Newtonian one for distances larger than roughly 1.3, 2.2 and
5 softening lengths, correspondingly for
the spline, p=4 and Plummer softening. 

In order to assess how well each softening can represent the forces in
a given mass distribution we calculated radial $MISE$ values for
different 
values of the softening, for different number of particles in the
configuration, and, in the case of the power softening, different
values of the exponent $p$, roughly in the range from 2 to 8. 
For all types of softening the $MISE$ as a function of $\epsilon$ curves
are very similar to those of Figure~\ref{repeat}, so we will not repeat
them here. 

Figure~\ref{eps_p_spline} compares the optimal softening
($\epsilon_{opt}$) for the power law softening - as a function of the
exponent $p$ - and for the spline softening. The values were obtained from
six hundred realisations of a Plummer sphere of 10~000 particles each,
two hundred realisations of 30~000 particles and sixty 
100~000 particle realisations. We note
that $\epsilon_{opt}$ increases with $p$.
Thus comparing $p$~=~2.0 to $p$~= 7.0 we find an increase
in $\epsilon_{opt}$ of roughly a factor of two. The $\epsilon_{opt}$
value for
the spline is somewhat larger than that of the highest $p$ values.
As expected, the optimal softening decreases with increasing number of
particles $N$. The $\Delta(log\epsilon_{opt})$ does not depend notably on
the power $p$. 

Figure~\ref{MISE_p_spline} compares the minimum value for radial $MISE$
($MISE_{opt}$) for the same cases as the previous figure. We note
that $MISE_{opt}$ decreases with $p$. Thus 
comparing $p$~=~2.0 to $p$~= 7.0 we find a 
decrease in $MISE_{opt}$ of roughly 30\%. The $MISE_{opt}$ value for
the spline is of the order of that of the highest $p$ values.
As expected, the corresponding minimum errors decrease with increasing
number of 
particles $N$. The $\Delta(logMISE_{opt})$ does not depend notably on
the power $p$. 

Figure~\ref{N_p_spline} compares the optimal softening and the
corresponding radial $MISE$ values as a function of $N$ for the spline
softening and for the power softening for exponents $p$ = 2 and $p$ = 5.
Power laws are satisfactory approximations in all cases, given by 

$$\epsilon_{opt}=AN^a$$

\noindent
and

$$MISE_{opt}=BN^b$$

\begin{table}
\centering
\caption{Coefficients of the power law fits }
\label{tab:spline_p}
\begin{tabular}{@{}ccccc@{}}
Method & A & a & B & b \\ 
$p$ = 2 & 0.79 & -0.25 & 0.36 & -0.72 \\
$p$ = 5 & 1.05 & -0.21 & 0.28 & -0.75 \\
spline & 1.22 & -0.20 & 0.28 & -0.75 \\
\end{tabular}
\end{table}

The values of the coefficients are given in
Table~\ref{tab:spline_p}. The small differences between the
coefficients for the Plummer sphere and Plummer softening given here
and those given in Table~\ref{tab:hpd} are due to the fact that here
we have used radial $MISE$ 
while in Table~\ref{tab:hpd} $MASE$. From this table, as 
well as from Figure~\ref{N_p_spline}, we see that the $MISE_{opt}$ as
a function of $N$ is more or less the same for the $p$ = 5 and the
spline. The $p$~=~2 case gives somewhat bigger values of
$MISE_{opt}$.  

All the above argue that the spline softening as well as the higher
values of the power in the power softening give a better
representation of the force than the standard Plummer softening. The
difference, however, is not as big as one could have inferred from   
Figure~\ref{compare_soft}, since some of the difference is compensated
by an adjustment in $\epsilon_{opt}$. Thus Figures~\ref{eps_p_spline}
and \ref{MISE_p_spline}
argue for an improvement of 30\%, with corresponding changes 
of $\epsilon_{opt}$ of a factor of two. This improvement is nevertheless
non-negligible, since it would take an increase of the number of
particles of roughly 70\% to achieve it
(cf. section~\ref{sec:plummer_opt}). The fact that the corresponding
value of $\epsilon_{opt}$ is higher is also an advantage, since, for
equally good representations of the forces in the mass distribution, a
larger softening allows for large time-steps, and therefore shorter CPU
execution times. Last but not least the spline softening necessitates
considerably less CPU time per call. This gain in time depends on whether
one programs in fortran or C, on what the exponent of
the power is, and on the compiler used. We have found ratios roughly
between 2 and 10. 

\section{Treecode}
\label{sec:tree}

By making some modifications to the standard treecode (Barnes and Hut
1986) it is possible to implement it on a GRAPE system (Makino 1991).
In particular the tree should not be descended for each
particle separately, but for blocks of particles, as initially
proposed by Barnes (1990). Increasing the number of particles in the
block makes the interaction list longer and the treecode 
more accurate. The particular implementation on the
Marseille GRAPE systems is described in A+98, together with some
discussion on its performance and accuracy. 

We have made calculations of $MASE$ using the GRAPE treecode and
radial $MISE$ with the standard one, for various values of the tolerance and
the number of particles $N$. The
differences between the values corresponding to the same number of
particles and different tolerances is rather small. In particular the
values obtained with a tolerance of 0.5 or 0.7 are very near those
obtained with
the direct summation. Only for tolerances larger than 1 do the
$MASE$ values increase significantly, and even so the differences with
the direct summation are always
considerably smaller than those obtained by changing the number of
particles by factors as those considered e.g. in Figure~\ref{repeat}. 

\section{Summary and Discussion}
\label{sec:discuss}
\indent
In this paper we have discussed the value of the softening that allows
us to best
approximate the true forces within a given mass distribution in an
N-body simulation. 

We have first worked with the Plummer sphere and confirmed previous
results that, for a given number of particles $N$, there is an optimal
softening which gives the best approximations to the forces. For
smaller values of the softening the noise introduces errors, while for
larger there is a systematic bias from the Newtonian force results. We
calculated the dependence of the optimal softening on the number of
particles $N$ and confirmed and extended the results of M96 and
A+98. We compared the results obtained for integrated and average
square errors and found that the latter were considerably less noisy,
as could be expected since they involve many more
samplings. For this reason $MASE$ calculations require considerably
more CPU time. Since we have at our disposal two powerful GRAPE systems
we performed $MASE$ calculations wherever this was possible, i.e. in
all cases with direct summation and the standard Plummer softening. We
used radial $MISE$ calculations for the other softenings. Some of the
treecode calculations were done on a GRAPE and some on a workstation.
We also find
that results obtained with GRAPE-3 agree very well with those obtained
on GRAPE-4, despite their differences in accuracy. This is due to the
fact that the errors in GRAPE-3 are 
due to round-off and can thus be considered as random (cf. A+98).

We then worked with other density distributions and found that the
density, and the central concentration, have a large influence on the
optimal softening and the corresponding errror $MASE_{opt}$. We first
examined the case of two concentric Plummer spheres of different
scale-length, and found that the existence of a dense sphere has a large
influence on the $\epsilon_{opt}$ and $MASE_{opt}$. We then compared
two other density distributions to the Plummer one, namely the
homogeneous and the Dehnen sphere. The former is much less centrally
concentrated and the latter much more than the Plummer sphere. 
This confirmed the importance of central concentration on the optimal
softening. All our results show that denser configurations necessitate
smaller softenings and are never as well represented as less dense
ones, i.e. they have considerably larger values of $MASE_{opt}$. 

Since the choice of the optimal softening depends on the configuration
under consideration, and it is rather cumbersome to perform a $MASE$
study for every different configuration, we propose a simple way of
obtaining a rough estimate of this optimal value. We show that it
depends on the mean distance of the nearest neighbours, a quantity which
is much easier to calculate or estimate than $MASE$. The precision of
this rough estimate should be sufficient in many cases.

We next examined the influence of the shape of the object, with the
help of Ferrers ellipsoids of different axial ratios. Although the
influence is large on the bias, it is very small for the optimal
softening and corresponding $MASE_{opt}$.

We then examined two alternative types of softening. One is a power-law
softening in which the value of the exponent can have values different
than 2, and the other is a spline softening. Large values of the
exponent as well as the spline necessitate a larger value of the
softening and give a smaller value of the $MASE_{opt}$. Since the
higher values of the exponent necessitate more CPU time than the
spline, it is the latter that provides the best ratio of
accuracy to CPU time. The difference, nevertheless, with the standard
Plummer softening is not very large.

The treecode results are somewhat less accurate than those obtained
with direct summation and the accuracy decreases with increasing
tolerance. The differences, however, are small. Thus in order
to obtain more accurate results within a given CPU time it should be 
preferable to increase the number of particles $N$,
rather than decrease the opening angle. This is particularly true for
the standard treecode, where the dependence of the CPU time on the
opening angle is stronger (Hernquist 1987, A+98). 

The question we have addressed in this paper is obviously of interest
for $N$-body simulations. If the adopted value of the softening is too
small, then the result will be very noisy. In $N$-body simulations one
considers only one realisation, thus the effect of noise can be very
acute.  
On the other hand if the adopted value of the softening is too large,
then the simulation will pertain to another object than was initially
desired. In certain cases this may be of little importance, provided
the softening is not excessive. This is not, however, always the case. For
example if we want to check the stability of a model obtained with the
Schwarzschild method (Schwarzschild 1979, 1982), then it is important that
the forces approximate as closely as possible those of the model.

The size of structures that need to be analysed has sometimes been
used in simulations to set the value of the softening. Our results,
however, show that this may not be always possible, and should
sometimes be accompanied by a corresponding increase in the number of
particles. Indeed if the size of the structures to be analysed are
smaller than the $\epsilon_{opt}$, then a better resolution
can be reached only by increasing the number of particles accordingly.

It is customary in N-body simulations to use a softening which is
constant both in time and position. Our results, however, show that
this practice should be questioned.

It is easy to implement a softening which is a function of
time. For example in the case we are simulating a collapse the
softening at the initial stages can be considerably larger than during
the stages of maximum collapse. This could also be said for head-on
encounters of galaxies, as e.g. in the formation of ring galaxies,
where the central concentration rises considerably over a short period
of time. Such a variation of the softening with time would be easy to
implement both on standard software and on GRAPE and could lead to
considerably more accuracy. 

More challenging is to consider a softening which a function of
position, as well as of time, particularly on GRAPE systems. 
For both GRAPE-3 and GRAPE-4 considerable modifications of the
software are necessary. The results of this paper, however, argue that
such an effort could be well worth-while. Such a work is in progress
(Athanassoula \& Lambert, in preparation). It should also be noted
that, when changing the softening, corresponding changes in the
time step should also be considered.

The fact that the optimal softening is a function of the number of
particles used in a simulation affects the CPU time necessary for a
given simulation. Suppose that we double the number of particles $N$
for a given simulation. Then
the simulation time will be multiplied by 4 if we are using direct
summation and roughly by 2 if we are using a treecode on GRAPE (A+98). However
the optimal 
softening will have also decreased by a factor which for a Plummer
type concentration will be of the order of 0.85. This will mean that
the time step should also be decreased by a similar factor, and the
total CPU time of the simulation increased accordingly. Thus the total
CPU time will increase roughly as $N^{2.24}$ for direct summation and as
$N^{1.24}$ for the treecode.

The dependence of the optimal softening on the number of particles
will also affect the relaxation time. Huang, Dubinski \& Carlberg
(1993) derived the following simple formula for the relaxation time as
a function 
of the softening and the number of particles, assuming that the
distribution of particles is homogeneous.

\begin{center}
\begin{equation}
T_{relax}={{N}\over{8ln(R/\epsilon)}}T_{cross}
\label{eq:trel_eps}
\end{equation}
\end{center}

\noindent
In the above $T_{cross}$ is the crossing time and $R$ is
some characteristic radius of the system.
Replacing the softening by its optimal value, which is a function of
the number of particles, we
find that the relaxation time will grow somewhat less rapidly then
expected if the softening was not a function of $N$.

These are the prices to pay for a simulation with better resolution,
with reduced particle noise  
and with more accurately calculated forces.

\parindent=0pt
\def\rr{\par\noindent\parshape=2 0cm 8cm 1cm 7cm}
\vskip 0.7cm plus .5cm minus .5cm

{\Large \bf Acknowledgments.} We would like to thank Yoko Funato, Dave Merritt,
Christos Vozikis and Mattias Wahde for useful discussions. We
would also like to thank the IGRAP, the
INSU/CNRS and the University of Aix-Marseille I for funds to develop
the GRAPE computing facilities used for the calculations in this paper.
\vskip 0.5cm

{\Large \bf References.}
\vskip 0.5cm
\parskip=0pt
\rr{Aarseth S. J. 1963 \MN, 126, 223}
\rr{Athanassoula E. 1993, in
{Probl\`emes \`a N corps et dynamique Gravitationnelle}, eds. F. 
Combes et E. Athanassoula, Editions de l'Observatoire de Paris}
\rr{Athanassoula E., Bosma A., Lambert J. C., Makino J, 1998, \MN,
293, 369 (A+98)}
\rr{Barnes J., 1990, J. Comput. Phys., 87, 161}
\rr{Barnes J. and Hut P. 1986, \Nat, 324, 446}
\rr{Barnes J. and Hut P. 1989, \ApJS, 70, 389}
\rr{Binney J. and Tremaine S. D. 1987, {Galactic Dynamics}, Princeton
University Press, Princeton}
\rr{Dehnen W. 1993, \MN, 265, 250}
\rr{Dyer C. C., Ip P. S. S. 1993 \ApJ, 409, 60}
\rr{Ferrers N.M. 1877, Q. J. Pure Appl. Math., 14, 1}
\rr{Hernquist L.: 1987, \ApJS, 64, 715}
\rr{Hernquist L., Hut P. and Makino J. 1993, \ApJL, 402, L85}
\rr{Hernquist L., Katz N., 1989 \ApJS, 70, 419}
\rr{Hockney R.W., Eastwood J.W.: 1981, 
``Computing simulations using particles", 
McGraw-Hill, New York}
\rr{Huang S., Dubinski J., Carlberg R.G. 1993, \ApJ, 404, 73}
\rr{Kawai A., Fukushige T., Taiji M., Makino J., Sugimoto D. 1997,
\PASJ, 49, 607}
\rr{Kawai A., Makino J. 1999 Proceedings of the Ninth SIAM Conference
on Parallel Processing for Scientific Computing, SIAM}
\rr{Makino J. 1991, \PASJ, 43, 621}
\rr{Makino J. 1994, in V.G. Gurza\-dyan, D. Pfenniger, eds, 
{Ergodic Concepts in Stellar Dynamics }, Springer Verlag, p. 131}
\rr{Makino J. Taiji M. 1998, {Scientific Simulations with
special-purpose computers}, John Wiley, Chichester}
\rr{Makino, J., Taiji, M., Ebisuzaki, T., Sugimoto, D. 1997, \ApJ, 480,
432}
\rr{Merritt D. 1996, \AJ, 111, 2462}
\rr{Palous J., Jungwiert B., Kopeck\'y J. 1993, \AAA, 274, 189}
\rr{Pfenniger D. Friedli D. 1993 \AAA, 270, 561}
\rr{Press W.H., Flannery B.P., Teukolsky S.A., Vetterling W.T. 1988 
Numerical Recipes in C, Cambridge Univ. press.
}
\rr{Salmon J.J., Warren, M.S.: 1994 J. Comp. Phys., 11, 136}
\rr{Schwarzschild M. 1979, \ApJ, 232, 236}
\rr{Schwarzschild M. 1982, \ApJ, 263, 599}
\rr{Sellwood J.A.: 1987, \AR, 25, 151}
\rr{Teuben P.J 1995 - in  R. Shaw, H.E. Payne and J.J.E. Hayes, eds, 
{Astronomical Data Analysis Software and
Systems IV}, ASP conference series 77, p. 398}

\label{lastpage}

\end{document}